# Delta Scuti Variables

Gerald Handler

*Institut für Astronomie, Türkenschanzstraße 17, 1180 Wien, Austria*

**Abstract.** We review recent research on Delta Scuti stars from an observer's viewpoint. First, some signposts helping to lead the way through the Delta Scuti jungle are placed. Then, some problems in studying individual pulsators in the framework of asteroseismology are given before a view on how the study of these variables has benefited (or not) from past and present high-precision asteroseismic space missions is presented. Some possible pitfalls in the analysis of data with a large dynamical range in pulsational amplitudes are pointed out, and a strategy to optimize the outcome of asteroseismic studies of Delta Scuti stars is suggested. We continue with some views on "hybrid" pulsators and interesting individual High Amplitude Delta Scuti stars, and then take a look on Delta Scuti stars in stellar systems of several different kinds. Recent results on pre-main sequence Delta Scuti stars are discussed as are questions related to the instability strip of these variables. Finally, some remarkable new theoretical results are highlighted before, instead of a set of classical conclusions, questions to be solved in the future, are raised.

**PACS:** Replace this text with PACS numbers; choose from this list: http://www.aip.org/pacs/index.html

## INTRODUCTION

The Delta Scuti stars are pulsating variables of spectral types A to F that lie at the intersection of the classical instability strip and the main sequence in the Hertzsprung-Russell (HR) Diagram (Fig. 1). Their oscillations are driven by the classical kappa mechanism in the HeII ionization zone. They have pulsation periods between about 18 minutes and 8 hours, corresponding to radial and nonradial p and mixed modes of low radial order. These pulsators comprise several different subtypes and behaviors that are worthwhile to be described to help finding some paths through the Delta Scuti jungle.

High-amplitude Delta Scuti stars (HADS) are defined as pulsators exceeding a visual light range of 0.3 magnitudes. In the HR Diagram, these variables lie within a narrow range at the center of the Delta Scuti instability strip. Most of these variables are (as known to date) pulsate in a dominant radial mode, usually the fundamental mode, and the majority of these single and double mode pulsators. However, a few triple-mode pulsators and stars that show simultaneous nonradial pulsation have been discovered as well. We will return to these objects later.

Whereas most Delta Scuti stars are Pop. I-type objects, there are also Pop. II stars. These are called SX-Phoenicis stars. They mostly occur in globular clusters and metal-poor galaxies, and are often conspicuous as blue stragglers, possibly originating

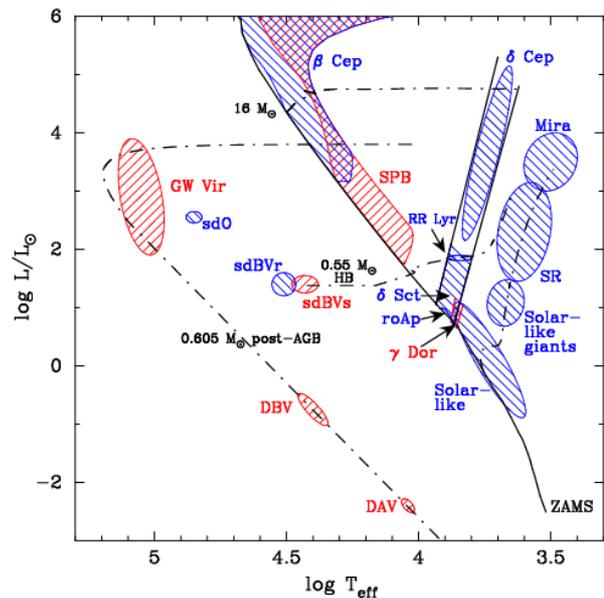

**FIGURE 1.** The Hertzsprung-Russell Diagram of pulsating stars. The Zero-Age Main Sequence is indicated as a full line, as is the classical instability strip (tilted from lower left to upper right). Some representative model evolutionary tracks are indicated as dash-dotted lines. Figure adapted and updated from [1].

from stellar mergers. Most SX Phe stars share the pulsational characteristics of the HADS.

At the effective temperatures of the Delta Scuti stars, diffusion and other processes can cause segregation of chemical elements. This modifies the excitation of the pulsations. From the point of view of theory, classical Am stars should not pulsate, but a few of them do (e.g., see [2]). The incidence of pulsation is higher in the Rho Puppis stars that are evolved Am stars, and the percentage of pulsators among the Lambda Bootis stars may even be higher than among the chemically "normal" Delta Scuti stars [3].

As pre-main sequence stars evolve towards the Zero-Age Main sequence, some of them cross the instability strip and may therefore also pulsate. These objects oscillate similar to the majority of the Delta Scuti stars, namely in multiple nonradial and possibly radial modes.

The final group to be mentioned here are oscillators in Algol-type eclipsing binaries, the oEA stars. These are often in semi-detached configurations, meaning that mass transfer is taking place. Some of the shortest-period Delta Scuti pulsators are located in such systems.

## ASTEROSEISMOLOGY OF DELTA SCUTI STARS AND THE SPACE PHOTOMETRY ERA

The multiperiodic oscillations of the Delta Scuti stars are of interest for the field of asteroseismology. This subfield of pulsating star research utilizes the individual stellar pulsations as seismic waves to explore the interior structures of stars by reproducing the observed pulsation modes via theoretical model calculations. The Delta Scuti stars are among the first pulsators whose asteroseismic potential has been recognized, yet their study remained notoriously difficult.

One of the reasons that makes the lives of Delta Scuti asteroseismologists hard is convection. These stars have shallow surface convection zones as well as convective cores. Material from the convective core can penetrate into the surrounding radiative regions, providing additional hydrogen for nuclear burning, and extending the stars' lifetimes on the main sequence. Asteroseismology is capable of yielding a measure of this convective overshooting [4], but for Delta Scuti stars this has not been done with certainty to date.

Surface convection, on the other hand, modifies the amplitude and phase of the bolometric flux with respect to a given radial velocity variation [5]. Initially seen as a problem, this has now turned into an asset for asteroseismology, as the parameters of the model used to describe surface convection can be constrained [6].

Another difficulty for asteroseismology of Delta Scuti stars is stellar rotation. Rapid rotation is difficult to model as it can no longer be treated with a perturbation approach and effects such as mode coupling provides further complications [7]. The logical consequence would be to study slowly rotating stars first. However, slowly rotating A/F stars often are Am or Ap stars that tend not to pulsate as Delta Scuti stars, whereas chemically normal stars with low projected rotational velocities may be intrinsic rapid rotators seen pole-on.

So far, Delta Scuti stars have also stubbornly refused to provide information on which modes they choose to pulsate in. Aside from a few systematics, such as hotter stars tend to pulsate in higher radial overtones [8], and that slowly rotating stars tend to pulsate with higher amplitudes and thus in radial modes [9], the problem of pulsational mode selection is not easy to solve. In addition, as Delta Scuti stars evolve, their convective cores shrink and their envelopes expand, causing a frequency increase of g modes trapped their interior, but a frequency decrease of p modes located in their envelope. At some point, when the frequencies of the two sets of modes become similar, the modes exchange physical properties and a dense spectrum of mixed modes results (e.g., see [10] for an illustration), severely hampering mode typing.

In this context, it is interesting to remember the development of observational studies of Delta Scuti pulsations for asteroseismic inferences. In the early days, measurements were taken from single sites during a few nights. This allowed the detection of a few (typically less than five) pulsation modes which is not sufficient to arrive at a unique seismic model (and can be harmed by aliasing problems!), but can be used to obtain series of possible models, in particular when combined with external constraints on the star's position in the HR Diagram.

Therefore, worldwide ground-based observing campaigns were initiated, typically yielding a few hundred hours of data and the frequencies of one or two dozen pulsation modes (e.g., see [11]). Still, these did not suffice for detailed comparisons with models as no constraints on the types of the excited pulsation modes aside from their frequencies were obtained.

Consequently, these efforts were again extended, obtaining of the order of one thousand hours of time-resolved multicolor photometry [12] and simultaneous spectroscopy [13] of individual stars for the purpose of identifying as many modes as possible with their spherical degrees and azimuthal orders.

At about the same time, first results from space observations of Delta Scuti stars became available (e.g., see [14]), but these did neither result in a significantly larger amount of detected pulsation modes, nor provide mode identifications.

However, dedicated asteroseismic space observations with the CoRoT satellite changed this picture [15]. Suddenly, more than a thousand possible pulsation modes were demonstrated to be excited in some stars. Now, after a quarter of a century of starving for more pulsation modes, we have detected too many to be modeled?!

Perhaps this surprising result should not be taken at face value. As has been pointed out before [16], one serious limitation of space photometry of opacity-driven main sequence pulsators is the limited time base most missions can supply. In observations with a time base of 100 d, only about 55 per cent of the theoretically predicted number of pulsation modes can be resolved. Prewhitening unresolved signals of high amplitude results in spurious small-amplitude peaks misleading frequency analyses.

The light curves of pulsating stars are not sinusoidal when working at high precision. Harmonic analysis of such signals results in combination frequencies. When working in a large dynamical range in amplitude, such as in the analysis of high-precision space photometry, some combination frequencies of the form $f_1 + f_2 - f_3$ can be present at a statistically significant level [17]. These must be identified with their "parent" modes and removed from the set of independent frequencies before a seismic analysis, a non-trivial task. The situation becomes even worse when higher-order combination frequencies are involved.

There are other circumstances that can be imagined as troublemakers for asteroseismology when analyzing high-quality data. Delta Scuti stars are known to show amplitude and frequencies variations at short time scales [11, 18]. (Sub)stellar companions also modulate oscillation frequencies through the light-time effect. In both cases, spurious signals are the result in straightforward frequency analyses.

Many of the Delta Scuti stars that have been, or will be, targets of asteroseismic space missions are faint. This raises the questions of how mode identifications can be obtained. Rather large ground-based telescopes are required to obtain sufficient signal-to-noise (both photometrically and spectroscopically), and time (base) on such telescopes is rare. A promising approach may be a combination of space photometry for high accuracy and ground-based observations for mode identification of bright stars [19]. For reasons of efficiency, the ground-based support would best done by a LCOGT [20] or SONG [21]-like network of automated telescopes.

### Delta Scuti Stars As "Hybrid" Pulsators

As Fig. 1 shows, the instability strip of the Delta Scuti stars overlaps with those of several other classes of variables, such as the RR Lyrae stars, the rapidly oscillating Ap (roAp) stars, the solar-like oscillators and the Gamma Doradus stars. Therefore it is conceivable that there are some objects that show the oscillations of more than one group of these variables. Indeed, [22] found Gamma Doradus and Delta Scuti oscillations in HD 209295 (CK Ind), but suggested at the same time that the g modes of the Gamma Doradus type may be tidally excited. The first example of an intrinsic Gamma Doradus/Delta Scuti "hybrid" is HD 8801 [23, 24, 25], for which two Delta Scuti pulsation frequencies and four or five Gamma Doradus modes were detected. Interestingly, three more oscillatory signals with pulsation constants exactly in between these two domains were also detected. Interestingly, HD 8801 appears to be a single star (and an Am star, [23]), suggesting that the intermediate periods may represent a previously unknown type of stellar pulsation in this part of the HR Diagram.

### SOME REMARKABLE HIGH AMPLITUDE DELTA SCUTI STARS

A multisite campaign of the SX Phe star BL Cam [26] revealed the presence of 22 independent pulsation modes, including by far the largest number of nonradial modes ever detected in a HADS. Periods down to 18 minutes were found, and the previously suspected double-mode nature of the star was rejected. Because of the faintness of the star, the low-amplitude pulsation modes could not be identified with the pulsational quantum numbers. Still, dedicated observing campaigns on HADS seem to be an enterprise worthwhile undertaking.

Systematic studies of HADS, such as [27] are therefore needed. Multiperiodic objects for follow-up projects can be pointed out, and pulsational period changes, possibly due to evolutionary effects or companions, can be measured.

Such efforts are carried out to an increasing extent by well-equipped amateur astronomers who make an effort to acquire large amounts of data. A remarkable work along these lines resulted in the discovery of only the fourth radial triple-mode pulsator among the Delta Scuti stars [28]. This object has the shortest period within its subclass and is therefore closest to the main sequence. On the other hand, the two triple-mode pulsators with the longest periods are well within the realm of the RR Lyrae stars [29].

Finally, the discovery of the first HADS in an eclipsing binary system is worth being mentioned [30].

# DELTA SCUTI STARS IN STELLAR SYSTEMS

Pop. I Delta Scuti stars are expected to be present in open clusters and have actually been searched for in these environments. This has several advantages for comparative studies, such as the realistic assumption that all these stars have the same metallicity and age, originating from the same interstellar cloud. From the point of view of asteroseismology, this means that modeling all stars in a given cluster must result in consistent external parameters to be credible. This approach is also called ensemble asteroseismology.

As an example, the open cluster NGC 6811 contains fourteen Delta Scuti stars [31]. This is particularly interesting, as this cluster lies in the field of view of the Kepler space mission (e.g., [32]) that is expected to provide high-quality photometry with a time base of three years. Another open cluster, NGC 6866 [33] is also located in the Kepler field, but only three Delta Scuti stars have been discovered in that cluster.

Another obvious astrophysical application of Delta Scuti stars is distance determination. These variables also follow a period-luminosity(-color) relation. For galactic Delta Scuti stars with accurate Hipparcos [34] parallaxes, [35] derived the following relation:

$$Mv = -2.90 \log P_F - 0.190 [Fe/H] - 1.27,$$

with an rms scatter of 0.16 mag; $P_F$ is the period of the fundamental radial mode. [36] used this relation, together with other means, to determine the distance modulus to the Large Magellanic Cloud (LMC) with 18.48 +/- 0.02 mag. This is consistent with the "long" distance scale to the LMC and with distances derived from Cepheids, RR Lyrae stars and from the red clump [37].

A systematic study of the Fornax dwarf spheroidal galaxy [38] led to the discovery of 85 HADS in this system. Interestingly, these objects comprised a significant number of subluminous stars, similar to the SX Phe stars in our Galaxy. However, the metal abundance in Fornax (-2.2 < [Fe/H] < -0.7) is well below that of the Milky Way, and therefore the criteria separating Delta Scuti and SX Phe stars in our Galaxy are not appropriate for the Fornax dwarf spheroidal [38]. In any case, these authors derived a period-luminosity relation from 183 HADS in extragalactic systems:

$$Mv = -3.65 \log P_F - 1.83,$$

with an rms scatter of 0.18 mag. The introduction of a color term in this equation does not make sense because the metallicities of these stars (the average V magnitude of the Delta Scuti stars in Fornax is about 23 mag) cannot be easily and well determined.

One may wonder why Delta Scuti stars would be used to determine distances to extragalactic systems at all because intrinsically much brighter standard candles would also be available. The answer is of practical nature: Delta Scuti stars have much shorter periods than, e.g., Cepheids, and therefore considerably less telescope time is necessary to derive sufficiently accurate periods.

# DELTA SCUTI STARS IN (ECLIPSING) BINARY SYSTEMS

One of the major problems in studying individual Delta Scuti stars is the determination of accurate stellar parameters such as their masses and luminosities. The available photometric and spectroscopic tools may not be free of systematic errors and therefore some more fundamental means are desirable.

A remarkable poster at this meeting [39] combined spectroscopic and astrometric constraints to derive accurate parameters of the visual binary $\Theta^2$ Tauri that probably consists of two Delta Scuti stars [40], including masses with a precision of 0.6 per cent.

At the time of this writing, about three dozen Delta Scuti stars in eclipsing binaries are known [41, 42, 43]. These would be very interesting objects to study asteroseismically, but again in practice this is more complicated than one would want. Most of these objects are rather faint and have low amplitudes, making mode detection difficult. As the probability of observable eclipses increases with decreasing orbital separation, many of the oEA stars have undergone or are undergoing mass transfer, which modifies their interior structure with respect to objects having evolved as single stars. As a consequence, even if such a pulsator can be studied seismically, the results would not be representative for the group.

On the other hand, an intriguing possibility remains: it has been shown that differential interior rotation modifies the apsidal motion rate of eccentric binary systems [44]. In turn, measuring the rate of apsidal motion, the amount of differential rotation can be inferred. The same can be done by applying asteroseismology. Therefore, it would be highly interesting to discover a multiperiodically pulsating component in an eccentric eclipsing binary system with rapid apsidal motion. In such a case, the inferences from both theories can be compared.

# PRE-MAIN SEQUENCE DELTA SCUTI STARS

Stars with initial masses larger than about 1.5 the solar value, evolving from the Hayashi track to the main sequence, cross the instability strip. Therefore, they are expected to show pulsational variability as well. The first pre-main sequence stars that show oscillations have been reported a long time ago [45], but the systematic study of this group has flourished only in recent years.

The main driver has been the recognition that pre-main sequence Delta Scuti stars are structurally simpler that their main sequence counterparts, which should also be reflected in their oscillation spectra. Therefore, these objects should be more easily treatable by theory.

Of course, Delta Scuti stars wouldn't be Delta Scuti stars wouldn't be there other complications. Aside from the fact that most pre-main sequence Delta Scuti stars are faint because they are in a relatively short-lived stage of evolution, they are also surrounded by circumstellar material that variably obscures the photosphere and modifies the observed pulsation amplitude and signal-to-noise of the data [46, 47]. To be fair, there are pre-main sequence Delta Scuti stars that do not misbehave and that have large numbers of pulsation frequencies excited [48].

As the pulsations of pre-main and main sequence Delta Scuti stars are driven by the same mechanism, it can be expected that their instability domains in the HR diagram are very similar. This has recently been observationally verified with success, although there is still a lack of accurate basic parameters of these stars due to missing parallaxes, temperature calibrations, because of circumstellar material and differential reddening within young open clusters [49].

# THE INSTABILITY STRIP OF DELTA SCUTI STARS

As one approaches the blue edge of the Delta Scuti domain, the pulsational driving region moves closer to the surface. At some point, there will no longer be sufficient resonating material to sustain observable pulsations. The blue edge of the Delta Scuti domain is therefore a direct consequence of the amount of Helium present in the driving zone, i.e. on the strength of driving. The red edge, on the other hand, is believed to be due to the interplay of convection and pulsational driving (e.g., [50]).

The currently accepted observational boundaries of the Delta Scuti instability strip [51] originate from the analysis of the latest catalogue of these variables [52]. This catalogue is based on measurements from the ground, but the resulting instability strip is nicely confirmed by recent CoRoT space photometry [53].

Another interesting finding from ground-based measurements is that the number of Delta Scuti stars increases considerably with decreasing amplitude [51], suggesting that in fact all stars within the Delta Scuti strip are pulsators if observed at sufficient accuracy. However, both the MOST [54] and CoRoT [55] satellites found constant stars within the Delta Scuti domain, down to a level of a few parts per million.

# THEORETICAL STUDIES OF DELTA SCUTI STARS

Radially pulsating Delta Scuti stars can yield a number of constraints on theoretical models. The period ratio between the radial fundamental and first overtone modes can be changed by varying the rotation rate or by using different opacities [56, 57, 58]. What is missing to this point is a detailed comparison with observed period ratios of radial multimode pulsators. In particular, new efforts to study radial triple mode pulsators are highly desirable.

The Delta Scuti star 44 Tau also has two radial modes excited [59, 60], plus a number of nonradial modes [59, 61]. Interestingly, the best-fitting seismic model for the star implies that we see 44 Tau in the rare post-main sequence contraction stage [62], additional impetus for detailed studies of this star.

For a few years, the puzzle of the chemical peculiarities of the Lambda Bootis stars seemed to be solved by the interpretation of the objects passing through interstellar gas clouds [63], modifying the superficial chemical composition. A recent seismic analysis of the pulsating Lambda Bootis star 29 Cyg [64] has revived this question, as the best-fitting models do not rule out this object as being submetallic throughout its whole interior.

Ensemble asteroseismology of Delta Scuti stars has also been an active research topic in the recent past. A new attempt to model five Delta Scuti stars in Praesepe has resulted in an internally consistent, and therefore encouraging, solution [65]. A similar attempt has been undertaken for pre-main sequence pulsators in NGC 6530 [66], but it remains to be seen how far the derived strong constraints on the basic parameters of the pulsators depend on the heavy assumptions in the modeling procedure.

## CONCLUSIONS

Due to the moderate requirements for time on large telescopes because of their short pulsation periods, the High Amplitude Delta Scuti stars prove to be useful distance indicators despite their low absolute magnitudes. It appears useful that these stars be exploited more by asteroseismic methods, as these are the (from a pulsational point of view) simplest oscillators. It seems unreasonable to expect understanding the most complicated (and thus astrophysically most interesting) pulsators before understanding the simple cases.

The "hybrid" pulsators also deserve special attention. At some point, it seemed that all of the Delta Scuti/Gamma Doradus "hybrids" (suggested to be named "Delgam Scudor stars" [67]) are Am stars [14]. Is this true, and if so, why would stars with an intrinsically low incidence of pulsation choose to oscillate in a more complicated way than others sharing the same parameter domain in the HR Diagram?

Many Delta Scuti stars show amplitude and frequency variations, but their simplest causes (like evolutionary changes or binarity) do everything but dominate the picture – what do these temporal variations in the oscillations, apparently driven by a robust mechanism, really mean?

A small fraction of stars in the Delta Scuti instability strip do apparently not pulsate. Is this a statistically sound observation? What physics separates the pulsators from the non-pulsators?

How can we finally make asteroseismic studies of Delta Scuti stars over a wide range of parameter space possible? Which stars may be the Rosetta Stones for asteroseismology? Well-studied, bright, high-amplitude, intrinsically slowly rotating stars? Double-mode radial pulsators with many additional nonradial modes? Components of eclipsing binary systems with accurately known masses and radii with multiperiodic oscillations? Will present and upcoming space missions give us measurements with the accuracy and time base needed to perform in-depth stellar structure modeling? Or will they provide information that we cannot understand at this point?

Whatever the answers to those questions are, one thing is sure: we will not run out of exciting questions concerning Delta Scuti stars in the near future.

## ACKNOWLEDGMENTS

I thank the organizers of this meeting for inviting me to present this review. This work has been supported by the Austrian Fonds zur Förderung der wissenschaftlichen Forschung under grant P20526-N16.

## REFERENCES

1. J. Christensen-Dalsgaard, "An Overview of Helio- and Asteroseismology", in *Helio- and Asteroseismology: Towards a Golden Future*, edited by D. Danesy, New Haven, Conneticut, 2004, ESA-SP 559, pp. 1 – 33
2. D. W. Kurtz, Mon. Not. Roy. Astron. Soc. **238**, 1077 – 1084 (1989)
3. E. Paunzen et al., Astron. Astrophys. **392**, 515 - 528 (2002)
4. W. A. Dziembowski and A. A. Pamyatnykh, Astron. Astrophys. **248**, L11 - L14 (1991)
5. L. A. Balona and E. A. Evers, Mon. Not. Roy. Astron. Soc. **302**, 349 – 361 (1999)
6. J. Daszynska-Daszkiewicz et al., Astron. Astrophys. **438**, 653 - 660 (2005)
7. A. A. Pamyatnykh, Ap. Sp. Sci. **284**, 97 – 107 (2003)
8. M. Breger and J. N. Bregman, Astrophys. J. **200**, 343 – 353 (1975)
9. M. Breger, Pub. Astr. Soc. Pacific, **94**, 845 - 849 (1982)
10. G. Handler, J. Astron. Astrophys., **26**, 241 - 247 (2005)
11. G. Handler et al., Mon. Not. Roy. Astron. Soc. **318**, 511 – 525 (2000)
12. M. Breger et al., Astron. Astrophys. **435**, 955 - 965 (2005)
13. W. Zima et al., Astron. Astrophys. **455**, 235 - 246 (2006)
14. J. M. Matthews, Comm. Asteroseism. 150, 333 – 340 (2007)
15. E. Poretti et al., these proceedings
16. G. Handler, "Beta Cephei stars as (asteroseismo)logical targets for EDDINGTON", in *Proceedings of the 2nd Eddington Workshop,* edited by F. Favata and S. Aigrain, ESA-SP 538, pp. 127 – 131
17. G. Handler et al., Mon. Not. Roy. Astron. Soc. **388**, 1444 – 1456 (2008)
18. M. Breger, these proceedings
19. H. Bruntt et al., Astron. Astrophys. **461**, 619 – 630 (2007)
20. T. Brown et al., Bull. Am. Astr. Soc., 41, 674 - 674
21. F. Grundahl et al., Comm. Asteroseism. 157, 273 – 278 (2008)
22. G. Handler et al., Mon. Not. Roy. Astron. Soc. **333**, 262 – 278 (2002)
23. G. W. Henry and F. C. Fekel, Astron. J., **129**, 2026 – 2033 (2005)
24. G. Handler, Mon. Not. Roy. Astron. Soc., in press (arXiv:0904.4859)
25. M.-P. Bouabid et al., these proceedings
26. E. Rodriguez et al., Astron. Astrophys. **471**, 255 – 264 (2007)
27. A. Derekas et al., Mon. Not. Roy. Astron. Soc. **394**, 995 – 1008 (2009)

28. P. Wils et al., Astron. Astrophys. **478**, 865 – 868 (2008)
29. G. Kovacs and R. Buchler, Astron. Astrophys. **281**, 749 – 755 (1994)
30. J. L. Christiansen et al., Mon. Not. Roy. Astron. Soc. **382**, 239 – 244 (2007)
31. Y. P. Luo et al., New Astronomy **14**, 584 – 589 (2009)
32. J. Christensen-Dalsgaard et al., Comm. Asteroseism. 150, 350 - 356 (2007)
33. J. Molenda-Zakowicz et al., Acta Astron. **59**, 193 – 211 (2009)
34. ESA, "The Hipparcos and Tycho Catalogues", 1997, ESA-SP 1200
35. D. H. McNamara et al., "The luminosities of horizontal branches and RR Lyrae stars in globular clusters", in *Variable Stars in the Local Group,* edited by D. W. Kurtz and K. R. Pollard, ASP Conf Ser. Vol. 310, pp. 525 - 529
36. D. H. McNamara et al., Astron. J. **133**, 2752 - 2763 (2007)
37. C. D. Laney and G. Pieterzynski, these proceedings
38. E. Poretti et al., Astrophys. J. **685**, 947 - 957 (2008)
39. P. Lampens et al., these proceedings
40. M. Breger et al., Astron. Astrophys. **336**, 249 - 258 (2002)
41. E. Soydugan et al., Mon. Not. Roy. Astron. Soc. **370**, 2013 – 2024 (2009)
42. A. Pigulski and G. Michalska, Acta Astron. **57**, 61 – 72 (2007)
43. D. Hoffman and T. Harrison, these proceedings
44. M. Yildiz, Astron. Astrophys. **409**, 689 - 695 (2003)
45. M. Breger, Astrophys. J. **171**, 539 – 548 (1972)
46. D. W. Kurtz and F. Marang, Mon. Not. Roy. Astron. Soc. **276**, 191 - 198 (1995)
47. K. Zwintz et al., Astron. Astrophys. **494**, 1031 - 1040 (2009)
48. T. Kallinger et al., Astron. Astrophys. **488**, 279 – 286 (2008)
49. K. Zwintz, Astrophys. J. **673**, 1088 - 1092 (2008)
50. M.-A. Dupret et al., Astron. Astrophys. **435**, 927 – 939 (2005)
51. E. Rodriguez and M. Breger, 2001, Astron. Astrophys. **366**, 178 – 196 (2001)
52. E. Rodriguez et al., Astron. Astrophys. Supp. Ser., **144**, 469 - 474 (2000)
53. A. Kaiser et al., these proceedings
54. G. A. H. Walker, talk given at the Second HELAS International Conference, Göttingen, Germany (2007)
55. L. Lefevre et al., "Blue Edge of the Delta Scuti Stars Versus Red Edge of the SPB Stars. How Will CoRoT Data Help?", in *Proceedings of the Annual meeting of the French Society of Astronomy and Astrophysics,* edited by C. Charbonnel et al., pp. 489 – 490
56. J. C. Suarez et al., Astron. Astrophys. **447**, 649 – 653 (2006)
57. J. C. Suarez et al., Astron. Astrophys. **474**, 961 – 967 (2007)
58. P. Lenz et al., "The Effect of Different Opacity Data and Chemical Element Mixture on the Petersen Diagram", in *Unsolved Problems in Stellar Physics: A Conference in Honor of Douglas Gough,* edited by R. J. Stancliffe et al., AIP Conf. Proc. Vol. 948, pp. 201 - 206
59. V. Antoci et al., Astron. Astrophys. **463**, 225 – 232 (2007)
60. P. Lenz et al., Astron. Astrophys. **478**, 855 – 863 (2008)
61. M. Breger and P.Lenz, Astron. Astrophys. **488**, 643 – 651 (2008)
62. P. Lenz et al., these proceedings
63. I. Kamp and E. Paunzen, Mon. Not. Roy. Astron. Soc. **335**, L45 - L49 (2002)
64. R. Casas et al., Astrophys. J., **697**, 522 – 534 (2009)
65. J. C. Suarez et al., Mon. Not. Roy. Astron. Soc. **379**, 201 – 208 (2007)
66. D. B. Guenther et al., Astrophys. J., **671**, 581 – 591 (2007)
67. R. R. Shobbrook, private communication